\documentclass[floats,floatfix,showpacs,preprintnumbers,amssymb,prd,twocolumn,superscriptaddress,nofootinbib,nolongbibliography,reprint]{revtex4-2}

\usepackage{amssymb,amsmath,verbatim,mathtools,needspace,enumitem,etoolbox,graphicx,physics,microtype,afterpage,xspace,tabularx,lmodern,multirow}
\usepackage{gensymb}
\usepackage[normalem]{ulem}
\usepackage[dvipsnames, usenames]{xcolor}
\usepackage{xr-hyper}
\definecolor{linkcolor}{rgb}{0.0,0.3,0.5}
\usepackage[unicode, colorlinks=true, linkcolor=linkcolor, citecolor=linkcolor, filecolor=linkcolor, urlcolor=linkcolor, linktocpage, breaklinks]{hyperref}
\usepackage[all]{hypcap}
\usepackage[T1]{fontenc}
\usepackage[utf8]{inputenc}
\usepackage[export]{adjustbox}
\usepackage[usenames,dvipsnames]{xcolor}
\hypersetup{colorlinks=true,citecolor=romared,linkcolor=romared,urlcolor=romared}

\definecolor{romared}{RGB}{142,0,28}

\newcommand{\be}{\begin{equation}}
\newcommand{\ee}{\end{equation}}

\def\be{\begin{equation}}
\def\ee{\end{equation}}
\newcommand{\beq}{\begin{eqnarray}}
\newcommand{\eeq}{\end{eqnarray}}

\usepackage{makecell}
\usepackage{soul}

\newcolumntype{Y}{>{\centering\arraybackslash}X}

\definecolor{romared}{RGB}{142,0,28}

\makeatletter
\newcommand*{\addFileDependency}[1]{
  \typeout{(#1)}
  \@addtofilelist{#1}
  \IfFileExists{#1}{}{\typeout{No file #1.}}
}
\makeatother

\begin{document}
\title{Can black holes be formed by focussing radiation?}

\author{Diego Blas} 
\affiliation{Institut de Fisica d'Altes Energies (IFAE), The Barcelona Institute of Science and Technology, Campus
UAB, 08193 Bellaterra (Barcelona), Spain}
\affiliation{Instituci\'o Catalana de Recerca i Estudis Avançats (ICREA), Passeig Llu\'{\i}s Companys 23, 08010 Barcelona, Spain}
\author{Vitor Cardoso} 
\affiliation{Niels Bohr Institute, Blegdamsvej 17, 2100 Copenhagen, Denmark}
\affiliation{CENTRA, Departamento de F\'{\i}sica, Instituto Superior T\'ecnico -- IST, Universidade de Lisboa -- UL, Avenida Rovisco Pais 1, 1049-001 Lisboa, Portugal}
\author{Jose Mar\'ia Ezquiaga} 
\affiliation{Niels Bohr Institute, Blegdamsvej 17, 2100 Copenhagen, Denmark}

\date{\today}

\begin{abstract}
Black holes are extreme outcomes of General Relativity, and can form through a variety of ways, including gravitational collapse of massive stars, or quantum fluctuations in the early universe. Here, we ask the question of whether they can form via focusing of radiation by compact binaries or intense lasers, or via trapping at the light ring of black holes.
We provide evidence that gravitational lensing of radiation from a small, finite number of sources is not a viable mechanism to form black holes.
\end{abstract}

\maketitle



\section {Introduction} 
\begin{figure}[h!]
\centering
\includegraphics[width = \columnwidth,valign=t]{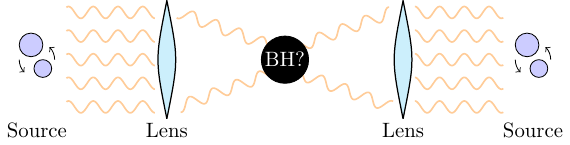}
\caption{Schematic diagram of a possible lensing system that focuses radiation from highly luminous event (the coalescence of a compact binary, say) to produce a black hole (BH).}
\label{fig:lensing_geometry}
\end{figure}
Black holes (BHs) are an essential solution, and an inescapable conclusion of stellar collapse, in General Relativity. Besides stars which ran out of nuclear fuel~\cite{PhysRev.56.455,PhysRev.141.1232}, alternative mechanisms to form BHs include trans-Planckian energy collisions or fluctuations in the primordial universe~\cite{Eardley:2002re,Zeldovich:1967lct,Hawking:1971ei}. Difficulties -- mostly related to pair creation -- in creating BHs from light have recently been studied~\cite{Alvarez-Dominguez:2024pub,Loeb:2024kui,Alvarez-Dominguez:2024mjd}, recovering a decades-old discussion on the existence and stability of geons~\cite{Wheeler:1955zz}. 

Here, we entertain an alternative production mechanism to form BHs: through the focusing and collapse of radiation. A possible mechanism to produce BHs (see Fig.~\ref{fig:lensing_geometry} for a depiction) might consist on a system of lenses (regardless of their nature) that would focus radiation to the point where it would collapse to a BH. For example,
take two weak beams of light, focus each appropriately (such that the power per unit area is sufficiently large) and collide them. Would this result in a BH? Alternatively, shine light onto a BH -- itself a gravitational lens -- and wait that radiation piles up at the light ring: will one get mini-BHs forming at the light ring? We wish to show here that neither of these outcomes is likely to occur, if the radiation is emitted by a finite and small\footnote{The term `small' will be characterized in the main text.} number of sources. 

\section{Can we focus radiation to produce a black hole?} 
We are concerned with possible BH formation through the lensing of gravitational, electromagnetic or any massless waves somewhere in the cosmos. We will use the Hoop Conjecture~\cite{1972mwm..book..231T,Choptuik:2009ww} (see also Ref.~\cite{Kaloper:2007pb} for an interesting discussion of BH formation from off-center collisions) as a discriminator for BH formation: when a system of mass $M$ (in $D=4$ dimensions) gets compacted into a region whose circumference in every direction has radius $\lesssim R=2GM/c^2$ , a horizon – and thus a BH – forms (here $G$ is Newton's constant and $c$ the speed of light in vacuum). In particular, take a {\it single} violent process producing a large amount of radiation. Can we lens it (by a BH or star or any human-made device), to focus radiation to such an extent that a BH is formed? A study of exact solutions of Einstein's equations describing impulsive waves, shows that parallel light beams don't attract~\cite{Faraoni:1998rc}. Hence, simply smoothly self-focusing a single beam of radiation is unlikely to produce collapse.
We therefore require two such beams to collide. The Hoop Conjecture applies well when the object being ``hooped'' is at rest (in this case the final black hole), so we consider two equal beams colliding head on. This scenario could describe, for instance, radiation from a binary coalescence being lensed by a supermassive BH or neutron star, and BH formation could occur at the focal point.

We consider a setup where the radiation from a binary coalescence, say, is focused by a system of lenses, and then made to overlap with a similar focused ``beam'' from another binary\footnote{Alternatively, radiation from a single source could be split in $n$ components, each of which can be focused and made to overlap with the others, in an attempt to produce BHs.}.
In other words, we assume that we are technologically sophisticated enough that we can capture the radiation emitted by a binary coalescence, emitted to all portions of the sky, and convert it into a nearly plane fronted ``beam'' of radiation. This beam will be assumed to be of high frequency\footnote{We make this choice to avoid having to deal with diffraction and wave effects in general. 
Diffraction of the incoming wave around the lens sets a maximum magnification $\mu_{\rm max}$ which grows linearly with its frequency $f$ and the mass of the lens $M_L$ \cite{1981Ap&SS..78..199B}, i.e. $\mu_{\rm max}\approx 4\times 10^{5}\left( M_L/M_\odot\right)\left(f/\mathrm{GHz}\right)$. 
Since higher magnifications are achieved in the high-frequency limit, this makes our argument more conservative.}, of cross-sectional area $A$ and intensity (energy per square meter per second) $I$. 
We assume that a system of lenses is placed properly to focus it to intensity $\tilde{I}$ and cross sectional area $\tilde{A}$. 
Energy conservation\footnote{Or, more specifically, conservation of phase-space volume (Liouville's theorem) in curved space-time along a ray bundle \cite{1973grav.book.....M,1992grle.book.....S}.} requires that
\be
I A=\tilde{I}\tilde{A}\,,\label{E_cons}
\ee
The mass density in such a beam is
\be
\rho=\frac{IA}{Ac}\times \frac{1}{c^2}=\frac{I}{c^3}.
\ee

Now we take one such beam and collide it with its equal. The overlapping region has length $\tilde{l}$ in all directions such that $\tilde{A}\sim \tilde{l}^2$. The Hoop Conjecture requires that, for a BH to form,
\be
{\cal C}\equiv \frac{2G}{c^2}\frac{M_{\rm inside}}{\tilde{l}}\gtrsim 1\,,
\ee
where $M_{\rm inside}$ is the mass within the cube of volume $\tilde{l}^3$,
\be
M_{\rm inside}=2\tilde{\rho}\tilde{l}^3\,,
\ee
therefore
\be
{\cal C}= \frac{4G}{c^5}\tilde{I} \tilde{l}^2\,,
\ee
which by \eqref{E_cons} is
\be
{\cal C}= \frac{4G}{c^5} I A\,,\label{eq:final}
\ee

Note that the conclusion above is not simply a statement about energy conservation. It depends crucially on the fact that the ``size'' of a four-dimensional BH scales with its mass (embodied in the Hoop Conjecture), and on the fact that the luminosity $IA$ is a dimensionless quantity in four-dimensions. The conclusion may change in higher dimensions, where the Hoop Conjecture takes on a different form~\cite{Ida:2002hg} and the maximum luminosity bound ceases to exist or to even make sense~\cite{Gibbons:2002iv,Jowsey:2021gny}.

In conclusion, if we want to form a BH via gravitational lensing of radiation, then the radiation being lensed should satisfy a bound on its luminosity ${\cal L}\equiv IA \gtrsim c^5/4G$. According to the maximum luminosity conjecture, this is impossible~\cite{Thorne:1982cv,Barrow:2014cga,Cardoso:2018nkg}. 
LIGO-Virgo-KAGRA events have ${\cal L}\sim 10^{-2}\,{c^5/\rm G}=4\times 10^{50}\,{\rm W}$ at its peak. It is certainly not feasible from the lensing of radiation from a BH binary. 

Note that the analysis above does not contradict in any way the findings of Christodoulou~\cite{Christodoulou:2008nj}, nor those of Choptuik~\cite{Choptuik:1992jv}. These authors find collapse to BHs from ``gravitational focusing'' since initial data in the latter is not constrained by any physics, and is possibly appropriate to collapsing scenarios. In fact, if we take $N$ sources (a good proxy for the spherically symmetric collapse in Ref.~\cite{Choptuik:1992jv}), estimate \eqref{eq:final} would be multiplied by $N$ and for sufficiently large $N$ the Hoop Conjecture would be satisfied and BHs formed. 

\section{Can we shine radiation onto a light ring and produce black holes?} 
The above does not take into account radiation re-cycling. In particular, Eq.~\eqref{E_cons} does not hold true in strong lensing, in trapping regions. We now turn our attention to regions where radiation is trapped and passes more than once. Light rays with an impact parameter
\be
b_{\rm crit}=3\sqrt{3}\frac{GM}{c^2}\,,
\ee
will orbit a non-spinning BH indefinitely, performing an infinite number of orbits around the circular null geodesic~\cite{Berti:2009bk,Berti:2010ce,Chandrasekhar:1985kt}. Imagine a situation where a non-spinning BH is being bombarded with radiation from some catastrophic event like the merger of two BHs. High frequency radiation hitting from one hemisphere orbits a large number ${\cal N}$ of times before it encounters radiation hitting from the opposite hemisphere head on.

For radiation to orbit a large number of times, it has to be fine-tuned close to the critical impact parameter, since ($\delta\equiv 1-b/b_{\rm crit}$)
\be
{\cal N}=\frac{-\log\delta}{\sqrt{2}\,\pi}\label{eq:critical}.
\ee
These trajectories get close to the light ring, at~\cite{Chandrasekhar:1985kt}
\be
r=3\frac{GM}{c^2}\left(1+\sqrt{2\delta/3}\right).
\ee

{\it Suppose} therefore that we can make radiation from a system ``fill'' the light ring and orbit it for a large number of times, say ${\cal N}$. Based on the above scaling, one might suspect that, {\it if} a BH forms, it could be sub-Planckian in size, and our classical estimates would not apply.

But it is easy to show that BHs can never form by filling the light ring. Take a binary of total mass $M_{\rm bin}$ close to a supermassive BH of mass $M$ which will lens the GWs. The total radiation emitted by a binary during its lifetime $E_{\rm rad}\lesssim 0.05M_{\rm bin}c^2$. It takes a time $\sim 30 GM/c^3$ for light to circle round the light ring, so even if all the radiation is channeled onto the light ring of the supermassive BH,
\be
{\cal L}\lesssim 10^{-6}\frac{10^3M_{\rm bin}}{M}\frac{c^5}{G}\,,
\ee
where we normalize to an optimistic value (for radiation to be high frequency the binary needs to be much smaller than the supermassive BH). Based on Eq.~\eqref{eq:final} with ${\cal L}=IA$ we conclude that we cannot form a BH this way. We thus rule out generic BH formation from lensing of gravitational waves from standard, {\it single} astrophysical sources.

What about bombarding a BH constantly with radiation, say the Cosmic Microwave Background (CMB)? Then effectively 
\be
{\cal C}= \frac{2G}{c^5} {\cal N} I A\,,\label{eq:final2}
\ee
where $A\sim G^2M^2\delta^2/c^4$ is the area of a surface centered at the critical impact parameter and ${\cal N}$ is given in \eqref{eq:critical}. Substituting,
\be
{\cal C}= -10^{-56}\delta^2\log\delta \frac{I}{10^{-9}\,{\rm W}/m^2} \frac{M}{M_{\odot}^2}\,.\label{eq:final3}
\ee
where we normalized to the CMB intensity~\cite{Winstein:2003zw}. Again, no BH formation is possible ($\delta\ll 1 $ by assumption). 

\section{Conclusions} 
Our results show how possible BH formation from ultra-intense lasers is just not a viable or very effective BH production mechanism, and that in general the production of trapped surfaces from focusing of radiation will not occur.
Our estimates apply to high-frequency radiation, for low frequencies similar results would apply as their larger wavelengths would prevent more focusing due to diffraction. 
These findings also echo similar results for fine-tuned initial data concerning high-energy collisions of two BHs~ \cite{Page:2022bem}, or results showing that BH formation is accompanied by extreme features

We have not included spin in our arguments, as it seems to worsen the situation: co- and counter-rotating particles would not even collide since the light rings are located at different radii.

Finally, we note that all our results and conclusions are based on the Hoop Conjecture. There are robust simulations supporting it~\cite{Choptuik:2009ww}, but it has nevertheless not been proven, nor formulated in a rigorous way.
as long as we focus on the examples considered in this paper. In addition, we investigated the possibility that radiation from isolated events could be focused so as to give rise to BH formation. We showed that this is not possible. However, we stress that there are other ways to produce black holes: critical collapse is an example of BH formation, which our universe could use to form primordial BHs. We argued that this can be understood as an extreme form of focusing from an infinite number of sources.

\noindent {\bf \em Acknowledgments.} 
%
We are thankful to Emanuele Berti, Tomohiro Harada, Masashi Kimura, Paolo Pani and Toni Riotto for useful correspondence.
We acknowledge support by VILLUM Foundation (grants no. VIL37766 and no.~VIL53101) and the DNRF Chair program (grant no. DNRF162) by the Danish National Research Foundation.
V.C.\ is a Villum Investigator and a DNRF Chair.  
V.C. acknowledges financial support provided under the European Union’s H2020 ERC Advanced Grant “Black holes: gravitational engines of discovery” grant agreement no. Gravitas–101052587. 
Views and opinions expressed are however those of the author only and do not necessarily reflect those of the European Union or the European Research Council. Neither the European Union nor the granting authority can be held responsible for them.
This project has received funding from the European Union's Horizon 2020 research and innovation programme under the Marie Sklodowska-Curie grant agreement No 101007855 and No 101131233. D. Blas acknowledges the support from the Departament de Recerca i Universitats from Generalitat de Catalunya to the Grup de Recerca 00649 (Codi: 2021 SGR 00649).
The research leading to these results has received funding from the Spanish Ministry of Science and Innovation (PID2020-115845GB-I00/AEI/10.13039/501100011033).
IFAE is partially funded by the CERCA program of the Generalitat de Catalunya.

\bibliography{References}

\begin{thebibliography}{29}%
\makeatletter
\providecommand \@ifxundefined [1]{%
 \@ifx{#1\undefined}
}%
\providecommand \@ifnum [1]{%
 \ifnum #1\expandafter \@firstoftwo
 \else \expandafter \@secondoftwo
 \fi
}%
\providecommand \@ifx [1]{%
 \ifx #1\expandafter \@firstoftwo
 \else \expandafter \@secondoftwo
 \fi
}%
\providecommand \natexlab [1]{#1}%
\providecommand \enquote  [1]{``#1''}%
\providecommand \bibnamefont  [1]{#1}%
\providecommand \bibfnamefont [1]{#1}%
\providecommand \citenamefont [1]{#1}%
\providecommand \href@noop [0]{\@secondoftwo}%
\providecommand \href [0]{\begingroup \@sanitize@url \@href}%
\providecommand \@href[1]{\@@startlink{#1}\@@href}%
\providecommand \@@href[1]{\endgroup#1\@@endlink}%
\providecommand \@sanitize@url [0]{\catcode `\\12\catcode `\$12\catcode
  `\&12\catcode `\#12\catcode `\^12\catcode `\_12\catcode `\%12\relax}%
\providecommand \@@startlink[1]{}%
\providecommand \@@endlink[0]{}%
\providecommand \url  [0]{\begingroup\@sanitize@url \@url }%
\providecommand \@url [1]{\endgroup\@href {#1}{\urlprefix }}%
\providecommand \urlprefix  [0]{URL }%
\providecommand \Eprint [0]{\href }%
\providecommand \doibase [0]{https://doi.org/}%
\providecommand \selectlanguage [0]{\@gobble}%
\providecommand \bibinfo  [0]{\@secondoftwo}%
\providecommand \bibfield  [0]{\@secondoftwo}%
\providecommand \translation [1]{[#1]}%
\providecommand \BibitemOpen [0]{}%
\providecommand \bibitemStop [0]{}%
\providecommand \bibitemNoStop [0]{.\EOS\space}%
\providecommand \EOS [0]{\spacefactor3000\relax}%
\providecommand \BibitemShut  [1]{\csname bibitem#1\endcsname}%
\let\auto@bib@innerbib\@empty
\bibitem [{\citenamefont {Oppenheimer}\ and\ \citenamefont
  {Snyder}(1939)}]{PhysRev.56.455}%
  \BibitemOpen
  \bibfield  {author} {\bibinfo {author} {\bibfnamefont {J.~R.}\ \bibnamefont
  {Oppenheimer}}\ and\ \bibinfo {author} {\bibfnamefont {H.}~\bibnamefont
  {Snyder}},\ }\href {https://doi.org/10.1103/PhysRev.56.455} {\bibfield
  {journal} {\bibinfo  {journal} {Phys. Rev.}\ }\textbf {\bibinfo {volume}
  {56}},\ \bibinfo {pages} {455} (\bibinfo {year} {1939})}\BibitemShut
  {NoStop}%
\bibitem [{\citenamefont {May}\ and\ \citenamefont
  {White}(1966)}]{PhysRev.141.1232}%
  \BibitemOpen
  \bibfield  {author} {\bibinfo {author} {\bibfnamefont {M.~M.}\ \bibnamefont
  {May}}\ and\ \bibinfo {author} {\bibfnamefont {R.~H.}\ \bibnamefont
  {White}},\ }\href {https://doi.org/10.1103/PhysRev.141.1232} {\bibfield
  {journal} {\bibinfo  {journal} {Phys. Rev.}\ }\textbf {\bibinfo {volume}
  {141}},\ \bibinfo {pages} {1232} (\bibinfo {year} {1966})}\BibitemShut
  {NoStop}%
\bibitem [{\citenamefont {Eardley}\ and\ \citenamefont
  {Giddings}(2002)}]{Eardley:2002re}%
  \BibitemOpen
  \bibfield  {author} {\bibinfo {author} {\bibfnamefont {D.~M.}\ \bibnamefont
  {Eardley}}\ and\ \bibinfo {author} {\bibfnamefont {S.~B.}\ \bibnamefont
  {Giddings}},\ }\href {https://doi.org/10.1103/PhysRevD.66.044011} {\bibfield
  {journal} {\bibinfo  {journal} {Phys. Rev. D}\ }\textbf {\bibinfo {volume}
  {66}},\ \bibinfo {pages} {044011} (\bibinfo {year} {2002})},\ \Eprint
  {https://arxiv.org/abs/gr-qc/0201034} {arXiv:gr-qc/0201034} \BibitemShut
  {NoStop}%
\bibitem [{\citenamefont {Zel'dovich}\ and\ \citenamefont
  {Novikov}(1967)}]{Zeldovich:1967lct}%
  \BibitemOpen
  \bibfield  {author} {\bibinfo {author} {\bibfnamefont {Y.~B.}\ \bibnamefont
  {Zel'dovich}}\ and\ \bibinfo {author} {\bibfnamefont {I.~D.}\ \bibnamefont
  {Novikov}},\ }\href@noop {} {\bibfield  {journal} {\bibinfo  {journal} {Sov.
  Astron.}\ }\textbf {\bibinfo {volume} {10}},\ \bibinfo {pages} {602}
  (\bibinfo {year} {1967})}\BibitemShut {NoStop}%
\bibitem [{\citenamefont {Hawking}(1971)}]{Hawking:1971ei}%
  \BibitemOpen
  \bibfield  {author} {\bibinfo {author} {\bibfnamefont {S.}~\bibnamefont
  {Hawking}},\ }\href {https://doi.org/10.1093/mnras/152.1.75} {\bibfield
  {journal} {\bibinfo  {journal} {Mon. Not. Roy. Astron. Soc.}\ }\textbf
  {\bibinfo {volume} {152}},\ \bibinfo {pages} {75} (\bibinfo {year}
  {1971})}\BibitemShut {NoStop}%
\bibitem [{\citenamefont {\'Alvarez-Dom\'\i{}nguez}\ \emph
  {et~al.}(2024{\natexlab{a}})\citenamefont {\'Alvarez-Dom\'\i{}nguez},
  \citenamefont {Garay}, \citenamefont {Mart\'\i{}n-Mart\'\i{}nez},\ and\
  \citenamefont {Polo-G\'omez}}]{Alvarez-Dominguez:2024pub}%
  \BibitemOpen
  \bibfield  {author} {\bibinfo {author} {\bibfnamefont {A.}~\bibnamefont
  {\'Alvarez-Dom\'\i{}nguez}}, \bibinfo {author} {\bibfnamefont {L.~J.}\
  \bibnamefont {Garay}}, \bibinfo {author} {\bibfnamefont {E.}~\bibnamefont
  {Mart\'\i{}n-Mart\'\i{}nez}},\ and\ \bibinfo {author} {\bibfnamefont
  {J.}~\bibnamefont {Polo-G\'omez}},\ }\href
  {https://doi.org/10.1103/PhysRevLett.133.041401} {\bibfield  {journal}
  {\bibinfo  {journal} {Phys. Rev. Lett.}\ }\textbf {\bibinfo {volume} {133}},\
  \bibinfo {pages} {041401} (\bibinfo {year} {2024}{\natexlab{a}})},\ \Eprint
  {https://arxiv.org/abs/2405.02389} {arXiv:2405.02389 [gr-qc]} \BibitemShut
  {NoStop}%
\bibitem [{\citenamefont {Loeb}(2024)}]{Loeb:2024kui}%
  \BibitemOpen
  \bibfield  {author} {\bibinfo {author} {\bibfnamefont {A.}~\bibnamefont
  {Loeb}},\ }\href@noop {} {\  (\bibinfo {year} {2024})},\ \Eprint
  {https://arxiv.org/abs/2408.06714} {arXiv:2408.06714 [gr-qc]} \BibitemShut
  {NoStop}%
\bibitem [{\citenamefont {\'Alvarez-Dom\'\i{}nguez}\ \emph
  {et~al.}(2024{\natexlab{b}})\citenamefont {\'Alvarez-Dom\'\i{}nguez},
  \citenamefont {Garay}, \citenamefont {Mart\'\i{}n-Mart\'\i{}nez},\ and\
  \citenamefont {Polo-G\'omez}}]{Alvarez-Dominguez:2024mjd}%
  \BibitemOpen
  \bibfield  {author} {\bibinfo {author} {\bibfnamefont {A.}~\bibnamefont
  {\'Alvarez-Dom\'\i{}nguez}}, \bibinfo {author} {\bibfnamefont {L.~J.}\
  \bibnamefont {Garay}}, \bibinfo {author} {\bibfnamefont {E.}~\bibnamefont
  {Mart\'\i{}n-Mart\'\i{}nez}},\ and\ \bibinfo {author} {\bibfnamefont
  {J.}~\bibnamefont {Polo-G\'omez}},\ }\href@noop {} {\  (\bibinfo {year}
  {2024}{\natexlab{b}})},\ \Eprint {https://arxiv.org/abs/2408.11097}
  {arXiv:2408.11097 [gr-qc]} \BibitemShut {NoStop}%
\bibitem [{\citenamefont {Wheeler}(1955)}]{Wheeler:1955zz}%
  \BibitemOpen
  \bibfield  {author} {\bibinfo {author} {\bibfnamefont {J.~A.}\ \bibnamefont
  {Wheeler}},\ }\href {https://doi.org/10.1103/PhysRev.97.511} {\bibfield
  {journal} {\bibinfo  {journal} {Phys. Rev.}\ }\textbf {\bibinfo {volume}
  {97}},\ \bibinfo {pages} {511} (\bibinfo {year} {1955})}\BibitemShut
  {NoStop}%
\bibitem [{\citenamefont {{Thorne}}(1972)}]{1972mwm..book..231T}%
  \BibitemOpen
  \bibfield  {author} {\bibinfo {author} {\bibfnamefont {K.~S.}\ \bibnamefont
  {{Thorne}}},\ }in\ \href@noop {} {\emph {\bibinfo {booktitle} {Magic Without
  Magic: John Archibald Wheeler}}},\ \bibinfo {editor} {edited by\ \bibinfo
  {editor} {\bibfnamefont {J.~R.}\ \bibnamefont {{Klauder}}}}\ (\bibinfo {year}
  {1972})\ p.\ \bibinfo {pages} {231}\BibitemShut {NoStop}%
\bibitem [{\citenamefont {Choptuik}\ and\ \citenamefont
  {Pretorius}(2010)}]{Choptuik:2009ww}%
  \BibitemOpen
  \bibfield  {author} {\bibinfo {author} {\bibfnamefont {M.~W.}\ \bibnamefont
  {Choptuik}}\ and\ \bibinfo {author} {\bibfnamefont {F.}~\bibnamefont
  {Pretorius}},\ }\href {https://doi.org/10.1103/PhysRevLett.104.111101}
  {\bibfield  {journal} {\bibinfo  {journal} {Phys. Rev. Lett.}\ }\textbf
  {\bibinfo {volume} {104}},\ \bibinfo {pages} {111101} (\bibinfo {year}
  {2010})},\ \Eprint {https://arxiv.org/abs/0908.1780} {arXiv:0908.1780
  [gr-qc]} \BibitemShut {NoStop}%
\bibitem [{\citenamefont {Kaloper}\ and\ \citenamefont
  {Terning}(2007)}]{Kaloper:2007pb}%
  \BibitemOpen
  \bibfield  {author} {\bibinfo {author} {\bibfnamefont {N.}~\bibnamefont
  {Kaloper}}\ and\ \bibinfo {author} {\bibfnamefont {J.}~\bibnamefont
  {Terning}},\ }\href {https://doi.org/10.1142/S0218271808012413} {\bibfield
  {journal} {\bibinfo  {journal} {Gen. Rel. Grav.}\ }\textbf {\bibinfo {volume}
  {39}},\ \bibinfo {pages} {1525} (\bibinfo {year} {2007})},\ \Eprint
  {https://arxiv.org/abs/0705.0408} {arXiv:0705.0408 [hep-th]} \BibitemShut
  {NoStop}%
\bibitem [{\citenamefont {Faraoni}\ and\ \citenamefont
  {Dumse}(1999)}]{Faraoni:1998rc}%
  \BibitemOpen
  \bibfield  {author} {\bibinfo {author} {\bibfnamefont {V.}~\bibnamefont
  {Faraoni}}\ and\ \bibinfo {author} {\bibfnamefont {R.~M.}\ \bibnamefont
  {Dumse}},\ }\href {https://doi.org/10.1023/A:1018867405133} {\bibfield
  {journal} {\bibinfo  {journal} {Gen. Rel. Grav.}\ }\textbf {\bibinfo {volume}
  {31}},\ \bibinfo {pages} {91} (\bibinfo {year} {1999})},\ \Eprint
  {https://arxiv.org/abs/gr-qc/9811052} {arXiv:gr-qc/9811052} \BibitemShut
  {NoStop}%
\bibitem [{\citenamefont {{Bontz}}\ and\ \citenamefont
  {{Haugan}}(1981)}]{1981Ap&SS..78..199B}%
  \BibitemOpen
  \bibfield  {author} {\bibinfo {author} {\bibfnamefont {R.~J.}\ \bibnamefont
  {{Bontz}}}\ and\ \bibinfo {author} {\bibfnamefont {M.~P.}\ \bibnamefont
  {{Haugan}}},\ }\href {https://doi.org/10.1007/BF00654034} {\bibfield
  {journal} {\bibinfo  {journal} {Astrophysics and Space Science}\ }\textbf
  {\bibinfo {volume} {78}},\ \bibinfo {pages} {199} (\bibinfo {year}
  {1981})}\BibitemShut {NoStop}%
\bibitem [{\citenamefont {{Misner}}\ \emph {et~al.}(1973)\citenamefont
  {{Misner}}, \citenamefont {{Thorne}},\ and\ \citenamefont
  {{Wheeler}}}]{1973grav.book.....M}%
  \BibitemOpen
  \bibfield  {author} {\bibinfo {author} {\bibfnamefont {C.~W.}\ \bibnamefont
  {{Misner}}}, \bibinfo {author} {\bibfnamefont {K.~S.}\ \bibnamefont
  {{Thorne}}},\ and\ \bibinfo {author} {\bibfnamefont {J.~A.}\ \bibnamefont
  {{Wheeler}}},\ }\href@noop {} {\emph {\bibinfo {title} {{Gravitation}}}}\
  (\bibinfo {year} {1973})\BibitemShut {NoStop}%
\bibitem [{\citenamefont {{Schneider}}\ \emph {et~al.}(1992)\citenamefont
  {{Schneider}}, \citenamefont {{Ehlers}},\ and\ \citenamefont
  {{Falco}}}]{1992grle.book.....S}%
  \BibitemOpen
  \bibfield  {author} {\bibinfo {author} {\bibfnamefont {P.}~\bibnamefont
  {{Schneider}}}, \bibinfo {author} {\bibfnamefont {J.}~\bibnamefont
  {{Ehlers}}},\ and\ \bibinfo {author} {\bibfnamefont {E.~E.}\ \bibnamefont
  {{Falco}}},\ }\href {https://doi.org/10.1007/978-3-662-03758-4} {\emph
  {\bibinfo {title} {{Gravitational Lenses}}}}\ (\bibinfo {year}
  {1992})\BibitemShut {NoStop}%
\bibitem [{\citenamefont {Ida}\ and\ \citenamefont {Nakao}(2002)}]{Ida:2002hg}%
  \BibitemOpen
  \bibfield  {author} {\bibinfo {author} {\bibfnamefont {D.}~\bibnamefont
  {Ida}}\ and\ \bibinfo {author} {\bibfnamefont {K.-i.}\ \bibnamefont
  {Nakao}},\ }\href {https://doi.org/10.1103/PhysRevD.66.064026} {\bibfield
  {journal} {\bibinfo  {journal} {Phys. Rev. D}\ }\textbf {\bibinfo {volume}
  {66}},\ \bibinfo {pages} {064026} (\bibinfo {year} {2002})},\ \Eprint
  {https://arxiv.org/abs/gr-qc/0204082} {arXiv:gr-qc/0204082} \BibitemShut
  {NoStop}%
\bibitem [{\citenamefont {Gibbons}(2002)}]{Gibbons:2002iv}%
  \BibitemOpen
  \bibfield  {author} {\bibinfo {author} {\bibfnamefont {G.~W.}\ \bibnamefont
  {Gibbons}},\ }\href {https://doi.org/10.1023/A:1022370717626} {\bibfield
  {journal} {\bibinfo  {journal} {Found. Phys.}\ }\textbf {\bibinfo {volume}
  {32}},\ \bibinfo {pages} {1891} (\bibinfo {year} {2002})},\ \Eprint
  {https://arxiv.org/abs/hep-th/0210109} {arXiv:hep-th/0210109} \BibitemShut
  {NoStop}%
\bibitem [{\citenamefont {Jowsey}\ and\ \citenamefont
  {Visser}(2021)}]{Jowsey:2021gny}%
  \BibitemOpen
  \bibfield  {author} {\bibinfo {author} {\bibfnamefont {A.}~\bibnamefont
  {Jowsey}}\ and\ \bibinfo {author} {\bibfnamefont {M.}~\bibnamefont
  {Visser}},\ }\href {https://doi.org/10.1142/S0218271821420268} {\bibfield
  {journal} {\bibinfo  {journal} {Int. J. Mod. Phys. D}\ }\textbf {\bibinfo
  {volume} {30}},\ \bibinfo {pages} {2142026} (\bibinfo {year} {2021})},\
  \Eprint {https://arxiv.org/abs/2105.06650} {arXiv:2105.06650 [gr-qc]}
  \BibitemShut {NoStop}%
\bibitem [{\citenamefont {Thorne}(1982)}]{Thorne:1982cv}%
  \BibitemOpen
  \bibfield  {author} {\bibinfo {author} {\bibfnamefont {K.~S.}\ \bibnamefont
  {Thorne}},\ }in\ \href@noop {} {\emph {\bibinfo {booktitle} {{Les Houches
  Summer School on Gravitational Radiation}}}}\ (\bibinfo {year}
  {1982})\BibitemShut {NoStop}%
\bibitem [{\citenamefont {Barrow}\ and\ \citenamefont
  {Gibbons}(2015)}]{Barrow:2014cga}%
  \BibitemOpen
  \bibfield  {author} {\bibinfo {author} {\bibfnamefont {J.~D.}\ \bibnamefont
  {Barrow}}\ and\ \bibinfo {author} {\bibfnamefont {G.~W.}\ \bibnamefont
  {Gibbons}},\ }\href {https://doi.org/10.1093/mnras/stu2378} {\bibfield
  {journal} {\bibinfo  {journal} {Mon. Not. Roy. Astron. Soc.}\ }\textbf
  {\bibinfo {volume} {446}},\ \bibinfo {pages} {3874} (\bibinfo {year}
  {2015})},\ \Eprint {https://arxiv.org/abs/1408.1820} {arXiv:1408.1820
  [gr-qc]} \BibitemShut {NoStop}%
\bibitem [{\citenamefont {Cardoso}\ \emph {et~al.}(2018)\citenamefont
  {Cardoso}, \citenamefont {Ikeda}, \citenamefont {Moore},\ and\ \citenamefont
  {Yoo}}]{Cardoso:2018nkg}%
  \BibitemOpen
  \bibfield  {author} {\bibinfo {author} {\bibfnamefont {V.}~\bibnamefont
  {Cardoso}}, \bibinfo {author} {\bibfnamefont {T.}~\bibnamefont {Ikeda}},
  \bibinfo {author} {\bibfnamefont {C.~J.}\ \bibnamefont {Moore}},\ and\
  \bibinfo {author} {\bibfnamefont {C.-M.}\ \bibnamefont {Yoo}},\ }\href
  {https://doi.org/10.1103/PhysRevD.97.084013} {\bibfield  {journal} {\bibinfo
  {journal} {Phys. Rev. D}\ }\textbf {\bibinfo {volume} {97}},\ \bibinfo
  {pages} {084013} (\bibinfo {year} {2018})},\ \Eprint
  {https://arxiv.org/abs/1803.03271} {arXiv:1803.03271 [gr-qc]} \BibitemShut
  {NoStop}%
\bibitem [{\citenamefont {Christodoulou}(2008)}]{Christodoulou:2008nj}%
  \BibitemOpen
  \bibfield  {author} {\bibinfo {author} {\bibfnamefont {D.}~\bibnamefont
  {Christodoulou}},\ }in\ \href {https://doi.org/10.1142/9789814374552_0002}
  {\emph {\bibinfo {booktitle} {{12th Marcel Grossmann Meeting on General
  Relativity}}}}\ (\bibinfo {year} {2008})\ pp.\ \bibinfo {pages} {24--34},\
  \Eprint {https://arxiv.org/abs/0805.3880} {arXiv:0805.3880 [gr-qc]}
  \BibitemShut {NoStop}%
\bibitem [{\citenamefont {Choptuik}(1993)}]{Choptuik:1992jv}%
  \BibitemOpen
  \bibfield  {author} {\bibinfo {author} {\bibfnamefont {M.~W.}\ \bibnamefont
  {Choptuik}},\ }\href {https://doi.org/10.1103/PhysRevLett.70.9} {\bibfield
  {journal} {\bibinfo  {journal} {Phys. Rev. Lett.}\ }\textbf {\bibinfo
  {volume} {70}},\ \bibinfo {pages} {9} (\bibinfo {year} {1993})}\BibitemShut
  {NoStop}%
\bibitem [{\citenamefont {Berti}\ \emph {et~al.}(2009)\citenamefont {Berti},
  \citenamefont {Cardoso}, \citenamefont {Gualtieri}, \citenamefont
  {Pretorius},\ and\ \citenamefont {Sperhake}}]{Berti:2009bk}%
  \BibitemOpen
  \bibfield  {author} {\bibinfo {author} {\bibfnamefont {E.}~\bibnamefont
  {Berti}}, \bibinfo {author} {\bibfnamefont {V.}~\bibnamefont {Cardoso}},
  \bibinfo {author} {\bibfnamefont {L.}~\bibnamefont {Gualtieri}}, \bibinfo
  {author} {\bibfnamefont {F.}~\bibnamefont {Pretorius}},\ and\ \bibinfo
  {author} {\bibfnamefont {U.}~\bibnamefont {Sperhake}},\ }\href
  {https://doi.org/10.1103/PhysRevLett.103.239001} {\bibfield  {journal}
  {\bibinfo  {journal} {Phys. Rev. Lett.}\ }\textbf {\bibinfo {volume} {103}},\
  \bibinfo {pages} {239001} (\bibinfo {year} {2009})},\ \Eprint
  {https://arxiv.org/abs/0911.2243} {arXiv:0911.2243 [gr-qc]} \BibitemShut
  {NoStop}%
\bibitem [{\citenamefont {Berti}\ \emph {et~al.}(2010)\citenamefont {Berti},
  \citenamefont {Cardoso}, \citenamefont {Hinderer}, \citenamefont {Lemos},
  \citenamefont {Pretorius}, \citenamefont {Sperhake},\ and\ \citenamefont
  {Yunes}}]{Berti:2010ce}%
  \BibitemOpen
  \bibfield  {author} {\bibinfo {author} {\bibfnamefont {E.}~\bibnamefont
  {Berti}}, \bibinfo {author} {\bibfnamefont {V.}~\bibnamefont {Cardoso}},
  \bibinfo {author} {\bibfnamefont {T.}~\bibnamefont {Hinderer}}, \bibinfo
  {author} {\bibfnamefont {M.}~\bibnamefont {Lemos}}, \bibinfo {author}
  {\bibfnamefont {F.}~\bibnamefont {Pretorius}}, \bibinfo {author}
  {\bibfnamefont {U.}~\bibnamefont {Sperhake}},\ and\ \bibinfo {author}
  {\bibfnamefont {N.}~\bibnamefont {Yunes}},\ }\href
  {https://doi.org/10.1103/PhysRevD.81.104048} {\bibfield  {journal} {\bibinfo
  {journal} {Phys. Rev. D}\ }\textbf {\bibinfo {volume} {81}},\ \bibinfo
  {pages} {104048} (\bibinfo {year} {2010})},\ \Eprint
  {https://arxiv.org/abs/1003.0812} {arXiv:1003.0812 [gr-qc]} \BibitemShut
  {NoStop}%
\bibitem [{\citenamefont {Chandrasekhar}(1985)}]{Chandrasekhar:1985kt}%
  \BibitemOpen
  \bibfield  {author} {\bibinfo {author} {\bibfnamefont {S.}~\bibnamefont
  {Chandrasekhar}},\ }\href@noop {} {\emph {\bibinfo {title} {{The mathematical
  theory of black holes}}}}\ (\bibinfo {year} {1985})\BibitemShut {NoStop}%
\bibitem [{\citenamefont {Winstein}(2003)}]{Winstein:2003zw}%
  \BibitemOpen
  \bibfield  {author} {\bibinfo {author} {\bibfnamefont {B.}~\bibnamefont
  {Winstein}},\ }\href@noop {} {\bibfield  {journal} {\bibinfo  {journal}
  {eConf}\ }\textbf {\bibinfo {volume} {C0307282}},\ \bibinfo {pages} {L04}
  (\bibinfo {year} {2003})}\BibitemShut {NoStop}%
\bibitem [{\citenamefont {Page}(2023)}]{Page:2022bem}%
  \BibitemOpen
  \bibfield  {author} {\bibinfo {author} {\bibfnamefont {D.~N.}\ \bibnamefont
  {Page}},\ }\href {https://doi.org/10.1103/PhysRevD.107.064057} {\bibfield
  {journal} {\bibinfo  {journal} {Phys. Rev. D}\ }\textbf {\bibinfo {volume}
  {107}},\ \bibinfo {pages} {064057} (\bibinfo {year} {2023})},\ \Eprint
  {https://arxiv.org/abs/2212.03890} {arXiv:2212.03890 [gr-qc]} \BibitemShut
  {NoStop}%
\end{thebibliography}%

\end{document}